# Low-frequency noise and tunnelling magnetoresistance in Fe(110)/MgO(111)/Fe(110) epitaxial magnetic tunnel junctions

R.Guerrero, F.G. Aliev [*], and R.Villar

Departamento de Física de la Materia Condensada, C-III, Universidad Autónoma de Madrid, Cantoblanco, 28049, Madrid, Spain

J. Hauch, M. Fraune, and G. Güntherodt

II. Physikalisches Institut, Rheinisch-Westfälische Technische Hochschule Aachen, D-52074 Aachen, Germany

K.Rott, H.Brückl, and G.Reiss

Facultät für Physik, Universiteit Bielefeld, 33501 Bielefeld, Germany

We report on tunnelling magnetoresistance (TMR), current-voltage (IV) characteristics and low-frequency noise in epitaxially grown Fe(110)/MgO(111)/Fe(110) magnetic tunnel junctions (MTJs) with dimensions from 2x2 to 20x20 $\mu m^2$. The evaluated MgO energy barrier (0.50±0.08 eV), the barrier width (13.1±0.5 Å) as well as the resistance times area product (7±1 M$\Omega\mu m^2$) show relatively small variation, confirming a high quality epitaxy and uniformity of all MTJs studied. The noise power, though exhibiting large variation, was observed to be roughly anti-correlated with the TMR. Surprisingly, for the largest junctions we observed a strong enhancement of the normalized low-frequency noise in the antiparallel magnetic configuration. This behaviour could be related to an interplay between the magnetic state and the local barrier defects structure of the epitaxial MTJs.

Since the first observation of large tunnelling magnetoresistance (TMR) at room temperature in magnetic tunnel junctions (MTJs) [1-3], these have been one of the highlight topics in magnetoelectronics. The study of new types of epitaxial MTJs [4,5] is of special importance after the very recent observation of a large TMR at room temperature in two different systems with epitaxial MgO(100) barriers [6,7]. The near to three-fold increase of TMR in comparison to previously reported record values for amorphous $Al_2O_3$ barriers [8] was described in terms of coherent tunnelling in the case of Fe/MgO(100)/Fe MTJs [9,10] and alternatively attributed to large spin polarization of CoFe ferromagnetic layers in CoFe/MgO(100)/CoFe MTJs with epitaxial MgO(100) layer [7].

Here we report on the search for a correlation between TMR, low frequency noise and the device dimensions in a novel type of epitaxial MTJs Fe(110)/MgO(111)/Fe(110), yet unexplored for the (111) orientation of the oxide layer, as well as for the (110) orientation of Fe. Previous measurements of the noise in polycrystalline MTJs have provided a variety of new information such as: (i) observation of 1/f noise, independent of the relative orientation of ferromagnetic layers and related mainly to structural defects within the barrier [11,12], and (ii) presence of a well defined increase of the magnetic noise just near the transition regions between ferromagnetic and antiferromagnetic alignments [13,14].

Our measurements have been carried out on 14 Fe(110)/MgO(111)/Fe(110) epitaxial MTJs grown on a single substrate, with 4 different sizes from 2x2 to 20x20 $\mu m^2$. All our Fe/MgO/Fe multilayers structures have been grown in a UHV system (base pressure $1 \times 10^{-10}$ mbar) by molecular beam epitaxy (MBE) using $Al_2O_3(11\text{-}20)$ substrates. Typically, 10 nm Mo buffer layers have been grown at T = 1000 K before the Fe deposition. The first 50 nm Fe electrode has been grown at room temperature with a subsequent annealing at T = 600 K for 30 min to improve the crystalline quality. The MgO(111) barriers have been grown at T = 293 K from bulk MgO placed in a crucible of the electron beam evaporator. The second Fe(110) electrode (5 nm) has been deposited subsequently at room temperature on the MgO(111) barrier. To prevent oxidation all samples were protected by a 5nm Au cap layer. More details about growth and characterisation of the MgO(111)/Fe(110) system may be found in [15]. Thin film structures were microstructured using a combination of optical and electron beam lithography and $Ar^+$ ion milling. A $SiO_2$ insulating layer was deposited to prevent shortcuts between the lower electrode and the upper contact layer. Figure 1 shows a transmission electron microscopy cross sectional image of a MTJ.

For the transport measurements we used silver paste connected gold wires to contact the samples to four terminals on a chip carrier. The resistance of the silver paste and gold wires was negligible in comparison to the resistance of the MTJs, and therefore did not influence appreciably the results. In order to obtain the first and the second derivatives of the current-voltage (IV) characteristics a modulated (f < 77 Hz) DC current was sent through shunt resistors $R_s$ (1 <$R_s$< 10 MΩ) to the junction while the magnitude and phase of the AC response was detected by a lock-in amplifier. The typical frequency range (3 < f <1000 Hz) of the voltage noise was measured with an auto-spectrum technique by using a two-step amplifier. The signal, amplified up to $10^6$ times, was sent to a SR-780 spectrum analyser to compute the power spectrum.

Figure 1a shows typical room temperature magnetoresistance curves for 4 samples with dimensions 2x2, 5x5, 10x10 and 20x20 $\mu m^2$. The TMR was defined as: $(R_{AP} - R_P)/R_P$ with P denoting parallel and AP antiparallel magnetisation states. Clearly, the coercive field of the hard Fe layers stays nearly size independent, while the switching of the soft layer starts even before the magnetic field inversion. We attribute this unusual behaviour to the fringing fields at the sample edges, which could be of importance for small surface/perimeter ratio.

Figure 1b shows TMR as a function of the resistance times area (RA) product. It has been previously suggested [16] that a strong variation of the RA product with area indicates the presence of pinholes. The rather small variation in the RA parameter (7 ±1 MOhm µm$^2$), which we obtain with an area change of order 100, indicates that the studied MTJs may be considered, from the transport point of view, as nominally nearly identical, and that pinholes are not the main factor affecting the TMR. Moreover, the trend of decreasing TMR with increasing RA product is opposite to what one would expect in case that the pinholes affected the TMR. The negative temperature coefficient in R(T) curves observed in our MTJs, according to [17], also supports this point. Statistical analysis of the MgO barrier characteristics determined from differential conductance as a function of applied voltage is shown in Figure 1c. The MgO barrier parameters such as thickness and height of the MTJs vs. area were evaluated by using a parabolic Brinkman fit [18] of the IV curves. Interestingly, the MgO barrier height has been found to be 0.50 ±0.08 eV when averaged over all the samples studied. This value is about a factor of 5 lower than the expected value of bulk MgO [4]. Similar recent observations reported for Fe(100)/MgO(100)/Fe(100) were explained by the oxygen vacancy impurity band forming inside the wide (about 3 eV) energy gap of MgO [6].

The average width of the barrier was estimated to be reduced to about 13.1±0.5 Å from the nominally grown 40Å. This reduction could not be explained by a possible 3D growth of the MgO barrier because if it occurred, it should have produced a strong increase of the MgO/Fe interface roughness and, as a consequence, a strong variation in the area conduction. This contradicts our data ( Fig.1), which indicate rather good uniformity of the obtained barrier parameters. In our view, the reduced effective barrier width could be due to a non-rectangular energy profile of the barrier. Some FeO intermixing at the interface could also reduce the barrier width [15].

The spectral noise density (represented as V/Hz$^{1/2}$ and averaged over the range where the noise follows a 1/f dependence) follows, as expected, a linear dependence on the applied current (see inset in Fig.2a). With suppressed 1/f noise at small current, frequency independent contributions, including thermal (Johnson) and shot noise, were observed at the highest frequencies (Fig.2a). We shall analyse here only the 1/f noise which depends on the interaction of electrons with defects inside the barrier [19]. The noise characteristics of the studied MTJs are found to be more scattered than the main electron transport characteristics (conductivity and barrier parameters). Figure 2b shows the Hooge parameter α as a function of the TMR. The parameter α is defined [19] as $\alpha(f) = \frac{As^2(f)f}{V^2}$, where A is the junction area, f is the frequency and V is the voltage applied to the junction. It has been evaluated for H = 0 Oe, and we have averaged the noise over the frequency range where a 1/f behaviour was observed. This parameter allows to compare samples with different sizes and resistances. Despite a rather large variation in the noise characteristics, there seems to be a general trend of lower noise with increasing TMR. A similar tendency was also reported for polycrystalline CoFe/Al-oxide/CoFe MTJs [14], although with a much larger variation over 4 orders of magnitude in the noise parameter for nominally identical samples.

A detailed investigation of the noise spectra by changing the magnetic field in steps of 1 Oe, when driving the MTJ from P to AP and back to P alignment, shows, that the magnetic field dependence of the noise may be qualitatively different for the studied samples compared to polycrystalline samples. Some MTJs showed a noise parameter nearly independent of the relative alignment of the magnetic layers, with the exception of a narrow transition region where some excess magnetic noise could be seen (Fig.3). Other MTJs, however, revealed the presence of unexpected additional noise in the antiparallel state. A rather weak change of the form and the value of this extra noise with magnetic field, which once appeared in the AP state, points to its nonmagnetic (i.e. barrier structure related) origin. On the other hand, the strong variation of the noise (Hooge parameter) between parallel and antiparallel configurations (Fig.3) indicates that the noise level should be linked to the magnetization direction of the soft Fe layer. In our view, the unusual enhancement of the noise in the AP state could be caused by magnetoelastic constriction / elongation of the soft Fe layer following the field inversion.

Interestingly, the field dependent noise has been found to be present in the largest samples (20x20 $\mu m^2$) and to be absent in the small ones (2x2 $\mu m^2$ and 5x5 $\mu m^2$), while only some of the samples with intermediate dimensions (10x10 $\mu m^2$) demonstrated this behaviour. This further supports the magnetostriction origin of the noise enhancement suggested above, because in smaller MTJs the stray fields change the sign of the coercive field, enhancing magnetization in the AP state and reducing therefore the influence of the magnetostriction. Some recent experiments indicate the possible importance of inelastic stress in the AP state of MTJs. Recently, magnetostriction of ultrathin layers of magnetic d-metals [20] was found to be anomalously enhanced. Moreover, MTJs incorporating magnetostrictive free layers (FeCoBSi) have been demonstrated to work as extremely sensitive strain sensors [21]. Here we suggest the inverse effect to occur, when rotation of the free layer induces magnetostriction, which could be detected in 1/f noise through changes in the defect structure inside the tunnelling barrier.

*In summary* an intensive study of TMR, barrier parameters and noise for epitaxial Fe(110)/MgO(111)/Fe(110) MTJs shows a relatively small variation in the transport parameters, accompanied by a pronounced dispersion in the low frequency noise. For the largest junctions the normalised noise was found to be very different between parallel and anti-parallel states. This novel feature in the 1/f noise, absent in polycrystalline MTJs, could be due to the difference in stress acting on the MgO barrier.

Authors acknowledge useful discussions with A.P.Levanyuk, S.S.P.Parkin and E.Tsymbal. The work has been supported in parts by Spanish MEC ( MAT2003-02600) and Spanish-German Integrated Action project (HA-2002-0015).

FIGURE CAPTIONS

**Figure 1 (a) Tunnelling magnetoresistance (TMR) as function of the dimension of the MTJs at 300K. (b) Dependence of TMR on the RA product (c) Dependence of estimated barrier height and width on junction area. Inset shows typical cross sectional TEM image of the Fe(110)/MgO(111)/Fe(110) MTJ.**

**Figure 2 (a) Spectral density of noise of 20x20 $\mu m^2$s Fe(110)/MgO(111)/Fe(110) MTJ with RA=7.8 M$\Omega$ $\mu m^2$ for different applied bias voltages. Inset shows typical dependence of the 1/f noise on the bias. The slope provides noise to signal ratio for 1Hz of $2*10^{-4}$ (b) Dependence of the Hooge parameter at H = 0 Oe on TMR.**

**Figure 3 Dependence of the Hooge parameter noise parameter on magnetic field for 3 samples with different sizes. For comparison a typical TMR curve (solid line) for a 20x20 $\mu m^2$ MTJ is also shown.**

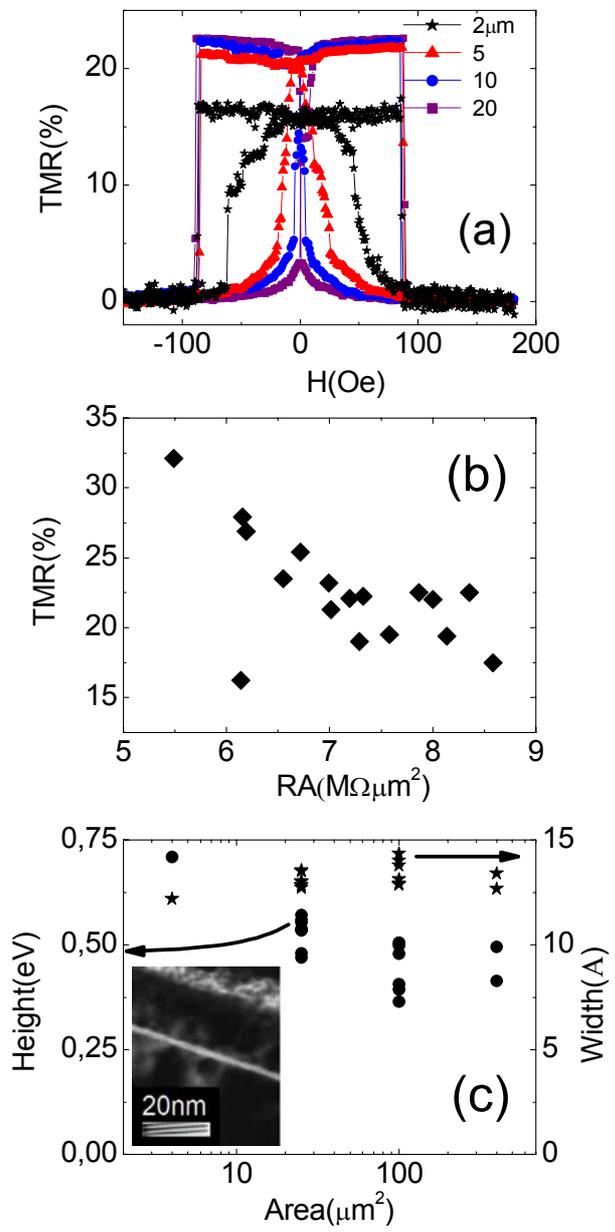

Fig. 1. Guerrero et al.

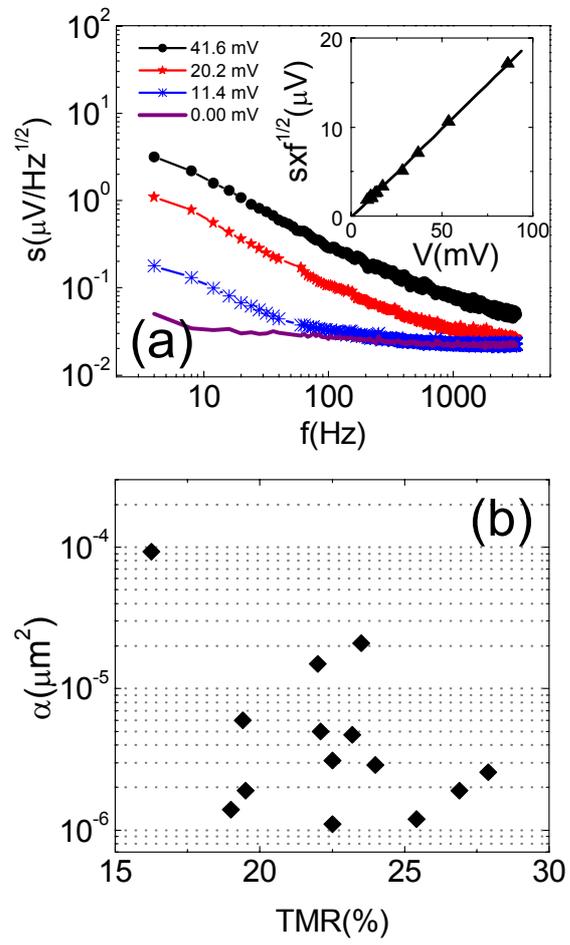

Fig. 2. Guerrero et al.

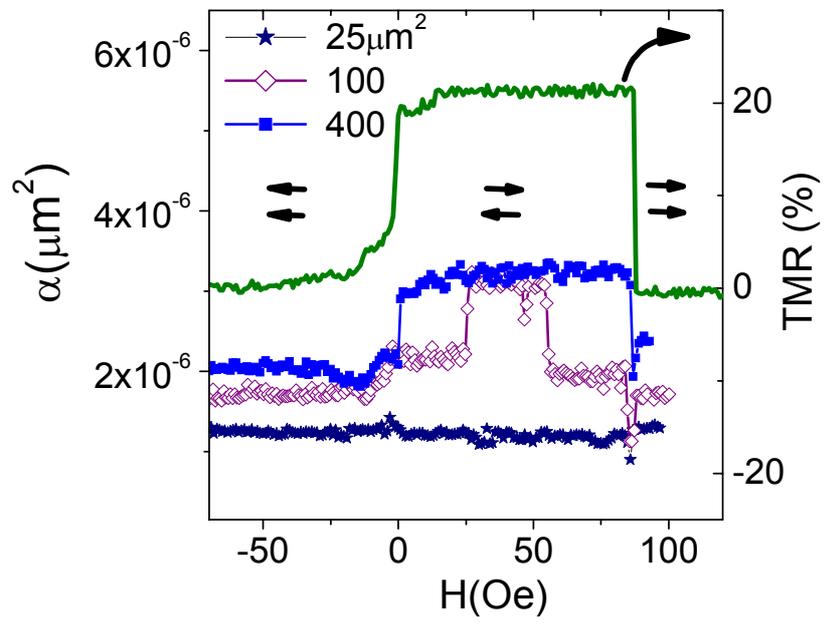

Fig. 3. Guerrero et al.